\documentstyle[12pt]{article}

\begin{document}
\thispagestyle{empty}
\begin{center}
\LARGE \tt \bf{Vacuumless torsion defects.}
\end{center}
\vspace{1cm}
\begin{center}
{\small \it
Departamento de Fisica Teorica-IF-UERJ-e-mail:garcia@dft.uerj.br.}\\
\end{center}
\vspace{0.5cm}
\begin{abstract}
Vacuumless defects in space-times with torsion may be obtained from vacuum defects in spacetimes without torsion.This idea is applied to planar domain walls and global monopoles.In the case of domain walls exponentially decaying Higgs type potentials are obtained.In the case of global monopoles torsion string type singularities are obtained like the string singularities in Dirac monopoles.
\end{abstract}
\vspace{0.5cm}
\begin{center}
\Large{PACS number(s) : 0420Jb, 0420Gz, 0450}
\end{center}
\newpage
\pagestyle{myheadings}
\markright{\underline{Vacuumless torsion defects}}
\section{Introduction}
\paragraph*{}
Recently Cho and Vilenkin have investigated the so-called vacuum defects which arises in symmetry breaking models when the scalar potential $V({\Phi})$ has no minima and is a monotonically decreasing function of the scalar potential.Strings,monopoles and domain walls have been obtained in this way.Here we provide two examples of vacuumless torsion defects where in the first case we show that vacuumless domain walls with torsion can be obtained from vacuum domain walls without torsion.The resultant domain walls are analogous to the supersymmetric domain walls obtained by Cvetic and Soleng \cite{1}.In the second case global monopoles with torsion are obtained from vacuumless global monopoles without torsion early given by Cho and Vilenkin.In this case torsion singularities of the string type are obtained in analogy of Dirac monopole.Both cases discussed here are treated from the point of view of the linear approximation of Einstein-Cartan theory of gravitation.In section 2 we deal with the domain wall case while in section 3 the global monopoles are handled.
\section{\bf Vacuumless torsionic domain walls.}
\paragraph*{}
Earlier M.Cvetic studied supersymmetric (extreme) domain walls in four-dimensional $N=1$ supergravity \cite{1}.In her paper a static conformally flat metric representing a wall linking Minkowski spacetime to a vacuum with a varying dilaton field was given.In some cases naked singularities could be formed on one side of the wall while in others the singularity of the wall could be covered by the horizon.Since as shown by Ferrara and Nieuwenhuizen \cite{2}that in supergravity torsion can be induced by gravitinos it would be very natural to investigate the analogy between supersymmetric domain walls and non-Riemannian domain wall solution of Einstein-Cartan gravity.In fact recently I have investigate non-Riemannian thin planar walls \cite{3,4} in Einstein-Cartan gravity, but I was not able to prove that those walls represented a planar domain wall since our system was a pure Einstein-Cartan gravity system and no scalar field was introduced.In this letter we in part solve this problem finding a solution of the complete system of the ECKG field equations.Since we want to show that a supersymmetric-type metric is a solution of the thick domain wall field equations we start from a metric of this type and simply check that it satisfies the ECKG field equations.Let us start from the generalization of a Riemannian thick domain wall discovered by Goetz \cite{5},and write down the ECKG field equations for the metric
\begin{equation}
ds^{2}=A(z)(dt^{2}-dz^{2}-b(t)(dx^{2}+dy^{2}))
\label{1}
\end{equation}
The energy-stress tensor is composed of two parts, first the scalar field 
${\Phi}$ with self-interactions contained in a potential $V({\Phi})$ given by 
\begin{equation}
{T_{ij}}^{Kink}= {\partial}_{i}{\Phi}{\partial}_{j}{\Phi}-g_{ij}(\frac{1}{2}{\partial}_{k}{\Phi}{\partial}^{k}{\Phi}-V({\Phi}))
\label{2}
\end{equation}
the other part is the torsion energy stress tensor given by
\begin{equation}
{{T_{i}}^{k}}^{torsion}={S_{iml}S^{kml}-\frac{1}{2}{{\delta}_{i}}^{k}{S_{0}}^{2}}
\label{3}
\end{equation}
where we have considered that the spin density tensor is totally skew symmetric as in the case 
of Dirac electrons and where $S_{ijk}$ is the torsion tensor and $S^{2}$ is the torsion-spin 
energy.The scalar field depends only on the z-coordinate which is orthogonal to the wall,i.e,we 
have ${\Phi}(z)$.The EC equation can be writen in the quasi-Einsteinian form
\begin{equation}
{G_{ij}}^{Riem}=kT_{ij}
\label{4}
\end{equation}
where $ T_{ij} $ is the sum of the stress-energy tensors given by {\ref{2}} and {\ref{3}} and 
${G_{ij}}^{Riem}$ is the Riemannian-Einstein tensor.The components of the stress-energy tensors 
are given by
\begin{equation}
{T^{t}}_{t}={T^{x}}_{x}={T^{y}}_{y}=e^{-{\lambda}}{{\Phi}'}^{2}+V({\Phi})-2{{S_{0}}^{2}}e^{-3{\lambda}}={\sigma}_{eff} 
\label{5}
\end{equation}
and 
\begin{equation}
{{T}_{z}}^{z}=-\frac{1}{2}(e^{-{\lambda}}{{\Phi}'}^{2}+V({\Phi})-3{S_{0}}^{2}e^{-3{\lambda}})= -p_{eff}
\label{6}
\end{equation}
where ${\sigma}_{eff}$ and $p_{eff}$ are respectively the effective 
density and pressure containing torsion energy and the Higgs type potential. 
The scalar field equation does not couple with torsion contrary to the point of view used in reference thus the Klein-Gordon field equation is the following
\begin{equation}
{{\Phi}"}+{{\Phi}'}({\psi}')=-\frac{dV}{d{\Phi}}e^{\lambda} 
\label{7}
\end{equation}
To simplify our task to find a solution we shall use the following ansatz
${\psi}'=S_{0}$ where $S_{0}$ is the constant torsion orthogonal to the wall.The Einstein-Cartan field equation reads
\begin{equation}
{G^{t}}_{t}-{G^{x}}_{x}=0
\label{8}  
\end{equation}
which implies the following equation
\begin{equation}
{b}^{..}b-{{b}^{.}}^{2}=0
\label{9}  
\end{equation}
The immediate solution of this equation is
\begin{equation}
b(t)=e^{ct}
\label{10}
\end{equation}
The other equations are 
\begin{equation}
-\frac{{{A}"}}{A^{2}}+\frac{3}{2}\frac{{A}'^{2}}{A^{3}}-\frac{c^{2}}{2A}=\frac{{{\Phi}'}^{2}}{A}
\label{11}
\end{equation}
and 
\begin{equation}
-\frac{{A'}^{2}}{A^{2}}+\frac{c^{2}}{A}=V(\Phi)-2\frac{{S_{0}}^{2}}{A^{3}}
\label{12}
\end{equation}
Integration of equation ${\psi}'=\frac{{A}'}{A}=S_{0}$ for constant torsion $S_{0}$ yields immediatly the solution $A(z)=e^{S_{0}z}$ which represents the dilatonic conformal factor of the supersymmetric metric.To simplify our computation we shall assume in the above equations that that the metric varies very slowly which means in the ultimate analysis that the torsion is very weak in second order and can be dropped which happens with the second derivatives of factor $A$.With this approximation and also considering that second derivatives of the scalar field vanish and the time factor $c$ vanishes since we are considering static metric a simple algebra yields the following results for the Higgs type potential and the energy $V({\Phi})$
\begin{equation}
V({\Phi})={S_{0}}^{2}(-{\alpha}+2e^{-3{\Phi}})
\label{13}
\end{equation}
and 
\begin{equation}
{\Phi}'={\alpha}\frac{A'}{A}
\label{14}
\end{equation}
or
\begin{equation}
{\Phi}={\alpha}{S_{0}}z+d
\label{15}
\end{equation}
where $d$ is an integration constant.Formula (\ref{15}) finally ends our proof.From the above expressions for $A$ we notice that the spacetime off the walls cannot be the Minkowskian unless torsion vanishes which cannot happen here for a constant torsion.Nevertheless this situation can be solved by introduction a torsion step function.In the present case the junction conditions for the ${\Phi}$ field yields
\begin{equation}
{\Phi}_{+}={\alpha}S_{0}{\epsilon}
\label{16}
\end{equation}
and
\begin{equation}
{\Phi}_{-}=-{\alpha}S_{0}{\epsilon}
\label{17}
\end{equation}
where $d=2{\epsilon}$ is the thickness of the domain wall which appears in the variation of the field across the wall given by 
\begin{equation}
{\Phi}_{+}-{\Phi}_{-}={\alpha}S_{0}d
\label{18}
\end{equation}
In this way torsion is shown to be explicitly responsible for the 
finite diference in the Higgs field in the same way matter is responsible 
for difference in the electric field across a pill-box in classical 
electrodynamics.Expression (\ref{18}) can be used to place an indirect limit on torsion.Although ${\epsilon}$ can be taken as small as we pleased 
it will never be zero due to physical constraint and therefore no real thin 
wall approximation could be possible otherwise we would not have a 
discontinuity on the Higgs type field as far as our model is concerned.
In the thin domain wall limit $T^{i}_{j}={\sigma}_{eff}diag(1,1,1,0)$ 
from expresions (\ref{4}) and (\ref{5}) we obtain the following surface 
energy density ${\sigma}_{eff}=\frac{1}{2}{S_{0}}^{2}({\alpha}-3e^{-3{\Phi}})$.
From a more recent paper by Jensen and Soleng \cite{6} we observe that the 
role played by the constant torsion here is analogous to the role played by 
the effective cosmological constant of the non-supersymmetric domain wall 
there.Further investigation on the analogies discussed here are now in 
progress.
\section{\bf Global torsionic monopoles.}
\paragraph*{}
Spacetime vacuumless defects \cite{7} have been recently investigated by 
Cho and Vilenkin.They have considered global and gauge defects like monopoles,
domain walls and strings in the approximation of Newtonian gravity and Einstein linearized approximation.Gauge monopole spacetime is essentially that of a magnetically charged black hole.Notice that since our main interest is provide an expression which allow us to place limits on torsion \cite{8} based on global monopoles parameters it is enough to use the Newtonian gravity approximation to obtain our results not taking into account other Riemannian terms of the metric.The investigation carried out here can also help to understand better the role played by torsion in the Early Universe.Let us now consider the following Newtonian gravity approximation
\begin{equation}
ds^{2}=-(1-2{\Phi})dt^{2}+(1+2{\Phi})dr^{2}+r^{2}d{\Omega}^{2}
\label{19}
\end{equation}
where $d{\Omega}^{2}= r^{2}(d{\theta}^{2}+sin^{2}{\theta}d{\phi}^{2})$ and ${\Phi}$ is the Newtonian potential.We assume here that this gravitational potential obeys what we called a Newton-Cartan field equations
\begin{equation}
{\nabla}^{2}{\Phi}=4{\pi}G(T^{0}_{0}-T^{i}_{i})
\label{20}
\end{equation}
where $T^{i}_{i}$ is the trace of the three-dimensional part of the energy-stress tensor which
is given by
\begin{equation}
T^{\nu}_{\mu}=({{\partial}_{\mu}}{\Phi}_{a})({{\partial}^{\mu}}{\Phi}_{a})-{\delta}^{\nu}_{\mu}L
\label{21}
\end{equation}
where L is the Lagrangean of the Higgs field.This expression will be added to the stress-energy tensor of torsion
\begin{equation}
{T_{\mu}^{\nu}}^{torsion}=S_{{\mu}{\alpha}{\beta}}S^{{\nu}{\alpha}{\beta}}-{\delta}_{\mu}^{\nu}{S_{0}}^{2}
\label{22}
\end{equation}
to yield the following extension of the Cho-Vilenkin equation for a static monopole
\begin{equation}
{\nabla}^{2}{\Phi}=-8{\pi}G(-\frac{2{\lambda}M^{4}}{a^{n}}(\frac{\delta}{r})^{\frac{2n}{n+2}}+S^{2}_{0})
\label{23}
\end{equation}
where ${\delta}$ is the size of the core defect and ${S_{0}}^{2}=S_{r{\theta}{\phi}}S^{r{\theta}{\phi}}=r^{-4}sin^{-2}{\theta}S^{2}_{r{\theta}{\phi}}$.
Where we have used the Minkowskian metric to raise and lower indices.To solve this equation we 
assume that the gravitational potential is a function only of the radial coordinate r.Therefore  
\begin{equation}
{\Phi}=\frac{2{\pi}(n+2)^{2}GM^{2}}{{a}^{n}(n+6)}(\frac{r}{\delta})^{\frac{4}{n+2}}-4{\pi}G{S_{0}}^{2}r^{2}
\label{24}
\end{equation}
Note that at equipotential gravitational surfaces defined by ${\Phi}=const.$,torsion can 
be easily determined.In particular in Minkowski space plus torsion torsion can be expressed as 
\begin{equation}
S^{2}_{r{\theta}{\phi}}=\frac{(n+2)^{2}GM^{2}}{2a^{n}(n+6)}(\frac{r}{\delta})^{\frac{-4(n+1)}{n+2}}sin^{-2}{\theta}
\label{25}
\end{equation}
From this expression we notice that when the observer is at distances much bigger that the 
radius of the core defect torsion vanishes unless $n=-1$, in this case torsion depends only on the angular direction and has a string type singularity at ${\theta}=0$ as in Dirac monopoles.Now let us consider the stress tensor given by the approximation of far zone as
\begin{equation}
T^{0}_{0}=T^{1}_{1}=\frac{{\eta}^{2}}{r^{2}}
\label{25}
\end{equation}
others zero.By considering that torsion is represented by the same stress tensor as above we 
obtain
\begin{equation}
{\Phi}(r)=8{\pi}G{\eta}^{2}lnr -4{\pi}GS^{2}_{0}r^{2}
\label{26}
\end{equation}
Substitution this expression into the above linearized metric we obtain the metric of the 
global torsionic monopole in the Newtonian approximation
\begin{equation}
ds^{2}=-(1-16{\pi}G{\eta}^{2}lnr+8{\pi}GS^{2}_{0}r^{2})dt^{2}+(1+16{\pi}G{\eta}^{2}lnr-8{\pi}GS^{2}_{0}r^{2})dr^{2}+r^{2}d{\Omega}^{2}
\label{27}
\end{equation}
It is interesting to note that in the absence of torsion this metric reduces to Barriola-Vilenkin spacetime when we expand the $lnr$ term.Note also that this spacetime does exert a gravitational force on the surrounding matter contrary to the Barriola-Vilenkin \cite{10} global monopole.This situation also happens in the case of Brans-Dicke global monopoles in linear gravity as investigated by Barros and Romero \cite{11}. To check explicitly this fact we compute the geodesic equation 
\begin{equation}
\frac{d^{2}x^{i}}{dt^{2}}=-\frac{1}{2}\frac{{\partial}h_{00}}{{\partial}x^{i}}
\label{28}
\end{equation}
we obtain the following expression
\begin{equation}
\frac{d^{2}r}{dt^{2}}=-\frac{{\partial}{\Phi}}{{\partial}r}
\label{29}
\end{equation}
Substitution the gravitational potential above we obtain
\begin{equation}
\frac{d^{2}r}{dt^{2}}=8{\pi}G({\eta}^{2}-2S^{2}_{0}r-{\frac{{\partial}S^{2}_{0}}{{\partial}r}}r^{2})
\label{30}
\end{equation}
By considering that the variation of torsion in the radial direction we conclude that although the nature of the gravitational field of the monopole is repulsive torsion can be used to slow down this repulsion and even when torsion is strong enough as in the early Universe one be able to turn the gravitational attractive field.Static external configurations can be obtained around the monopole by an appropriate choice of torsion.This can be used in gauge monopoles which are essentially magnetically charged blach holes to avoid its collapse.Finnally by comparison of metric (\ref{18}) with De Sitter metric 
\begin{equation}
ds^{2}=-(1-{\Lambda}r^{2})dt^{2}+(1+{\Lambda}r^{2})dr^{2}+r^{2}d{\Omega}^{2}
\label{31}
\end{equation}
If we compare this metric with our global torsionic metric we notice that the role played by torsion here is the same as played by the cosmological constant there.Other types of vacuumless defects as cosmic strings and textures can be investigated elsewhere.
\section*{Acknowledgments}
I am very much indebt to R.Ramos,A.Wang,P.S.Letelier,I.Shapiro,F.W.Hehl and H.Soleng for helpful discussions on the subject of this paper.Financial support from Universidade do Estado do Rio de Janeiro (UERJ) and FAPERJ is grateful acknowledged.

\end{document}